\documentclass[12pt,preprintnumbers,fullsize]{article}
\advance\hoffset by -1.2truecm
\advance\voffset by -1.5truecm
\usepackage{graphicx}
\usepackage{epsfig}
\usepackage{dcolumn}
\usepackage{bm}

\def\gsim{\lower0.5ex\hbox{$\:\buildrel >\over\sim\:$}}
\def\lsim{\lower0.5ex\hbox{$\:\buildrel <\over\sim\:$}}

\newcommand{\be}{\begin{equation}}
\newcommand{\ee}{\end{equation}}
\newcommand{\beq}{\begin{eqnarray}}
\newcommand{\eeq}{\end{eqnarray}}

\def\bra{\langle}
\def\ket{\rangle}

\def\a{\alpha}
\def\b{\beta}

\def\G{\Gamma}
\def\d{\delta}
\def\D{\Delta}
\def\e{\epsilon}

\def\m{\mu}
\def\n{\nu}
\def\p{\pi}
\def\ph{\phi}
\def\r{\rho}
\def\t{\tau}
\def\s{\sigma}

\begin{document}

\begin{flushright}
{TIFR/TH/04-13}\\
{June 2004}
\end{flushright}
\vspace{2cm}

\begin{center}
{\Large{\bf Looking for the Charged Higgs Boson}}
\vspace{2cm}

{\large{\bf D. P. Roy}} \vspace{5mm}


{\bf Department of Theoretical Physics} \\
{\bf Tata Institute of Fundamental Research} \\
{\bf Homi Bhabha Road, Mumbai 400 005, India} \vspace{3mm}

{dproy@theory.tifr.res.in}\par\vspace*{3cm}
\end{center}

\begin{abstract}
{This review article starts with a brief introduction to the charged
Higgs boson $(H^\pm)$ in the Minimal Supersymmetric Standard Model
(MSSM). It then discusses the prospects of a relatively light $H^\pm$
boson search via top quark decay at Tevatron/LHC, and finally a heavy
$H^\pm$ boson search at LHC. The viable channels for $H^\pm$
search are identified in both the cases, with particular emphasis on
the $H^\pm \to \t \n$ decay channel. The effects of NLO QCD correction in
the SM as well as the MSSM are discussed briefly.}
\end{abstract}


\section{Introduction}

The minimal supersymmetric extension of the Standard Model (MSSM)
contains two Higgs doublets $\ph^{+,0}_u$ and $\ph^{0,-}_d$, with
opposite hypercharge $Y=\pm 1$, to give masses to the up and down type
quarks and leptons. This also ensures anomaly cancellation between
their fermionic partners. The two doublets of complex scalars
correspond to 8 degrees of freedom, 3 of which are absorbed as
Goldstone bosons to give mass and longitudinal components to the
$W^\pm$ and $Z$ bosons. This leaves 5 physical states: two neutral
scalars $h^0$ and $H^0$, a pseudo-scalar $A^0$, and a pair of charged
Higgs bosons $H^\pm$. While it may be hard to distinguish any one of
these neutral Higgs bosons from that of the Standard Model, the
$H^\pm$ pair carry a distinctive hall-mark of the MSSM. Hence the
charged Higgs boson plays a very important role in the search of the
SUSY Higgs sector.

\subsection{Masses and Couplings} 

At the tree-level all the MSSM Higgs masses and couplings are given in
terms of two parameters -- the ratio of the vacuum expectation values,
$\tan \b = \bra \ph^0_u \ket / \bra \ph^0_d \ket$, and any one of the
masses, usually taken to be $M_A$. The physical $H^\pm$ and $A^0$
states correspond to the combinations
\beq
H^\pm &=& \ph^\pm_u \cos \b + \ph^\pm_d \sin \b , \nonumber \\ [2mm]
A^0 &=& \sqrt{2} ({\rm Im} \ph^0_u \cos \b + {\rm Im} \ph^0_d \sin \b) ,
\label{eq1}
\eeq
while their masses are related by
\be
M^2_{H^\pm} = M^2_A + M^2_W ,
\label{eq2}
\ee
with negligible radiative corrections \cite{one}. However, the neutral scalars
get a large radiative correction from the top quark loop along with
the top squark loop,
\be
\e = {3g^2 m^4_t \over 8 \p^2 M^2_W} \ell n \left( {M^2_{\tilde t} 
\over m_t^2} \right) , 
\label{eq3}
\ee
where $M_{\tilde t}$ denotes the average mass of the two top squarks
($\tilde t_{1,2}$). Including this radiative correction, the
mass-squared matrix of the neutral scalars is given by
\be
\pmatrix{M^2_A \sin^2\b + M^2_Z \cos^2 \b & -(M^2_A + M^2_Z) \sin \b \cos
\b \cr - (M^2_A + M^2_Z) \sin \b \cos \b & M^2_A \cos^2 \b + M^2_Z \sin^2
\b + \e'} ,
\label{eq4}
\ee
where $\e' = \e/\sin^2\b$. Thus the physical $h^0$ and $H^0$ masses
correspond to the eigen values
{\small
\be
M^2_{h,H} = {1\over 2} \left[M^2_A + M^2_Z + \e' \mp \left\{ (M^2_A +
M^2_Z + \e')^2 - 4 M^2_A M^2_Z \cos^2 \b - 4 \e' (M^2_A \sin^2 \b +
M^2_Z \cos^2 \b) \right\}^{1/2}\right].
\label{eq5}
\ee
}
The corresponding eigen vectors are the two orthogonal combinations of
${\rm Re}\ph^0_{u,d}$ with mixing angle $\a$, which diagonalizes this
matrix, i.e.
\be
\tan 2 \a = \tan 2 \b {M^2_A + M^2_Z \over M^2_A - M^2_Z + \e'/\cos 2 \b} , 
-\p/2 < \a < 0 .
\label{eq6}
\ee
For $M_A \gg M_Z, \a \to \b - \p/2$. Note that $M^2_H$ and
$M^2_{H^\pm} \to M^2_A$, while the lighter scalar mass approaches a
finite limit
\be
M^2_h {\buildrel {M_A \gg M_Z} \over \longrightarrow}
M^2_Z \cos^2 2 \b + \e .
\label{eq7}
\ee  
Finally there is an additional radiative contribution to this limit from
$\tilde t_{L,R}$ mixing \cite{one,two},
\be
\e_{\rm mix} = {3g^2 m^4_t \over 8\p^2 M^2_W} {X^2_t \over M^2_{\tilde t}} 
\left(1 - {X^2_t \over 12 M^2_{\tilde t}}\right) \leq {9g^2 m^4_t \over 
8\p^2 M^2_W} , 
\label{eq8}
\ee
where $X_t = A_t - \m \cot \b$. Thus while $\e_{\rm mix}$ is a
function of the SUSY breaking trilinear coupling $A_t$ and the
Higgsino mass parameter $\m$, it has a constant upper limit ($\sim M^2_W$),
which is reached at $X^2_t = 6 M^2_{\tilde t}$. One can also check
from (\ref{eq3}) that $\e \sim M^2_W$ for a SUSY breaking scale of
$M_{\tilde t} \sim $ 1 TeV. Adding the nonleading radiative
contributions to eqs.(\ref{eq7}) and (\ref{eq8}) gives a limit on the
light scalar mass
\be
M_h {\buildrel {M_A \gg M_Z} \over \longrightarrow} 118  {\rm GeV} (130
 {\rm GeV})  {\rm at} \tan \b = 3 (30) 
\label{eq9}
\ee
for the top quark pole mass of 175 GeV \cite{one, two}. 

\begin{table}
\caption{Important couplings of the MSSM neutral Higgs bosons $h$,
$H$ and $A$ relative to those of the SM Higgs boson.}
\[
\begin{tabular}{|c|c|c|c|c|}
\hline
&&&& \\
Channel & $H_{\rm SM}$ & $h$ & $H$ & $A$ \\
&&&& \\
\hline
&&&& \\
$\bar bb(\tau^+\tau^-)$ & $\displaystyle{gm_b \over 2M_W} (m_\tau)$ & $-\sin
\alpha/\cos \beta$ & $\cos \alpha/\cos \beta$ & $\tan \beta$ \\
 & & $\rightarrow 1$ & $\tan \beta$ & '' \\
&&&& \\
\hline
&&&& \\
$\bar tt$ & $\displaystyle g{m_t \over 2M_W}$ & $\cos\alpha/\sin\beta$ &
$\sin\alpha/\sin\beta$ & $\cot \beta$ \\
& & $\rightarrow 1$ & $\cot\beta$ & '' \\
&&&& \\
\hline
&&&& \\
$WW (ZZ)$ & $g M_W (M_Z)$ & $\sin (\beta - \alpha)$ & $\cos (\beta -
\alpha)$ & $0$ \\
& & $\rightarrow 1$ & $0$ & '' \\
&&&& \\
\hline
\end{tabular}
\]
\end{table}
Table 1 shows the important couplings of the MSSM neutral Higgs bosons
relative to those of the SM Higgs boson.  The limiting values of these
couplings at large $M_A$ are indicated by arrows.  The important 
couplings of the charged Higgs boson, which has no SM analogue, are 
\beq
H^+ \bar{t}b &:& {g \over \sqrt{2}M_W} (m_t \cot \b + m_b \tan
\b), \ H^+ \t \n : {g \over \sqrt{2}M_W} m_\t \tan \b, \nonumber \\[2mm]
H^+ \bar{c}s &:& {g \over \sqrt{2} M_W} (m_c \cot \b + m_s \tan \b), 
 H^+ W^- Z : 0 ,
\label{eq10}
\eeq
with negligible radiative corrections. 

The coefficients of the fermion mass terms of eq.(\ref{eq10}) and
Table 1 reflect the compositions of the respective Higgs bosons in
terms of $\ph_{u,d}$. It is clear from eqs.(\ref{eq2}) and
(\ref{eq10}) that measurements of $H^\pm$ mass and couplings will
determine the masses and couplings of the other MSSM Higgs bosons via
the underlying parameters $M_A$ and $\tan \b$.

\subsection{Indirect constraints on $\tan \b$ and $M_A$} 

The $H^+ \bar{t}b$ Yukawa coupling of eq.(\ref{eq10}) is ultraviolet
divergent. Assuming it to remain perturbative upto the GUT scale
implies
\be
1 < \tan \b < m_t/m_b (\sim 50) .
\label{eq11}
\ee
However this assumes the absence of any new physics beyond the MSSM
upto the GUT scale -- i.e. the socalled desert scenario. Without this
assumption one gets weaker limits from the perturbative bounds on this
coupling at the electroweak scale, i.e.
\be
0.3 < \tan \b < 200 .
\label{eq12}
\ee
Moreover there is a strong constraint on the $M_A - \tan \b$ parameter
space coming from the LEP-2 bound on the $H_{SM}$ mass, which is also
applicable to $M_h$ at low $\tan \b$, i.e. $M_h > 114$ GeV
\cite{two}. Comparing this with the MSSM prediction (\ref{eq9})
implies $\tan \b > 2.4$ for any value of $M_A$ \cite{one,two} (see
Fig. 4 below). Note however from eqs. (\ref{eq3}),(\ref{eq7}) and
(\ref{eq8}) that the MSSM prediction depends sensitively on the top quark
mass. The recent increase of this mass from 175 to 178 $\pm$ 4.3 GeV
\cite{three} along with a more exact evaluation of the radiative
correction \cite{four} have resulted in a significant weaking of this
constraint. In fact there is no LEP bound on $\tan \b$ now, which
would be valid for all values of $M_A$. Nonetheless it implies $M_A >$
150 GeV ($M_{H^\pm} >$ 170 GeV) over the low $\tan \b (\leq 2)$
region. But being an indirect bound, it depends strongly on the
underlying model. There is no such bound in the CP violating MSSM due
to $h$-$A$ mixing \cite{five}. Moreover there are singlet extensions
of the MSSM Higgs sector like the socalled NMSSM, which invalidate
these $M_A (M_{H^\pm})$ bounds without disturbing the charged Higgs
boson \cite{six}. In fact there is an additional contribution to the
tree-level mass relation (\ref{eq2}) in the NMSSM, which permits
$H^\pm$ to be even lighter than the $W$ boson. Therefore it is prudent
to relax these indirect constraints on $M_{H^\pm}$ and $\tan \b$, and
search for $H^\pm$ over the widest possible parameter space. It should
be noted here that the $H^\pm$ couplings of eq.(\ref{eq10}) continue
to hold over a wide class of models. In fact the fermionic couplings
hold for the general class of Type-II two-Higgs-doublet models, where
one doublet couples to up type and the other to down type quarks and
leptons \cite{one}.

\subsection{Direct $H^\pm$ Mass Limit from LEP}

Figure 1 shows a direct mass limit of $M_{H^\pm} > $ 80 GeV from
LEP-2, which is in agreement with the MSSM prediction (\ref{eq2}). It
is based on both the decay channels $H^\pm \to cs$ and $\t\n$ of
eq.(\ref{eq10}). Hence it is a robust limit, spanning the full $\tan
\b$ range of eq.(\ref{eq12}). The limit is broadly restricted to the
$W$ mass region because of the $W^+ W^-$ background. However one gets
a slightly stronger limit ($>$ 90 GeV) from the $\t\n$ channel, which
is reflected in the $\tan \b > 1$ region.

\section{Search for a Light $H^\pm (M_{H^\pm} < m_t)$ at Tevatron/LHC}

The main production mechanism in this case is top quark pair production 
\be
q\bar q, gg \to t\bar t ,
\label{eq13}
\ee
followed by
\be
t \to b H^+ \ {\rm and/or} \ \bar t \to \bar b H^- . 
\label{eq14}
\ee
The dominant decay channels of $H^\pm$ are 
\be
H^+ \to c \bar s, \t^+ \n \ {\rm and} \ W b \bar b + hc , 
\label{eq15}
\ee
where the 3-body final state comes via the virtual $t \bar b$
channel. All these decay widths are easily calculated from the Yukawa
couplings of eq.(\ref{eq10}). The QCD correction can be simply
implemented in the leading log approximation by substituting the quark
masses appearing in the Yukawa couplings by their running masses at
the $H^\pm$ mass scale \cite{seven}. Its main effect is to reduce the
$b$ and $c$ pole masses of 4.6 and 1.8 GeV respectively \cite{two} to
their running masses $m_b (M_{H^\pm}) \simeq $ 2.8 GeV and $m_c
(M_{H^\pm}) \simeq 1$ GeV. The corresponding reduction in the $t$ pole
mass of 175 GeV is only $\sim$ 5 \%.

The resulting branching ratios for the four decay processes of
(\ref{eq14}) and (\ref{eq15}) are shown in Fig. 2 against $\tan \b$
for a representative $H^\pm$ mass of 140 GeV. The $t \to bH^+$
branching ratio is seen to be large at $\tan \b \lsim 1$ and $\tan \b
\gsim m_t/m_b$, which are driven by the $m_t$ and the $m_b$ terms of
the $H^+ \bar t b$ coupling respectively. However it has a pronounced
minimum around $\tan \b = \sqrt{m_t/m_b} \simeq 7.5$, where the SM
decay of $t \to bW$ is dominant. The $H^\pm$ is expected to decay
dominantly into the $\t\n$ channel for $\tan \b > 1$, while the $cs$
and the $b\bar b W$ channels dominate in the $\tan \b \leq 1$
region. This can be easily understood in terms of the respective
couplings of eq.(\ref{eq10}). Note however that the $H^+ \to \bar b
bW$ three-body decay via virtual $t\bar b$ channel is larger than the
$H^+ \to c\bar s$ decay for $M_{H^\pm} \gsim $ 140 GeV, although the
former is a higher order process \cite{eight,nine}. This is because
the $H^+\bar t b$ coupling is larger than the $H^+\bar c s$ coupling
by a factor of $m_t/m_c > 100$ in the low $\tan \b$ region.
 
\subsection{$H^\pm$ Search in the $cs$ and $b\bar b W$ Channels 
($\tan \b < 1$)} 

One can look for a possible top quark decay into the $H^\pm \to cs$
channel in the Tevatron $t\bar t$ data in the leptonic and dileptonic
channels using the so called indirect or disappearance method
\cite{ten}. Here
\be
\s^\ell_{t\bar t} = \s_{t \dot{\bar t}} 2 B_\ell (1 - B_\ell), \ 
\s^{\ell\ell}_{t\bar t} = \s_{t\dot{\bar t}} B^2_\ell .
\label{eq16}
\ee
Using the QCD prediction for $\s_{t\bar t} = 5 - 5.5 pb$ \cite{eleven}
and SM prediction for $B_\ell \equiv B (t \to (e,\m) \n b) = 2/9$ one
can predict the number of such events. In the presence of $t \to bH (H
\to cs)$ decay channel one expect a reduction in $B_\ell$ and hence
the number of $t\bar t$ events in the leptonic and dileptonic channels
with respect to the SM prediction.

No such reduction was found in the $t\bar t$ data of the $D0\!\!\!/$
\cite{twelve} and CDF \cite{thirteen} experiments. The resulting exclusion
region in the $M_{H^\pm} - \tan \b$ parameter space is shown on the
left side of Fig. 1. It is seen to exclude the $M_{H^\pm} < 130$ GeV,
$\tan \b < 1$ region, where the $H^\pm \to cs$ is the dominant decay
mode. Indeed a comparison with Fig. 2 shows that the bulk of the
parameter space for which $H^\pm \to cs$ is the dominant decay mode is
already excluded by these data. This method is no longer applicable
for $M_{H^\pm} \geq 140$ GeV, where $H^\pm \to b\bar b W$ is the
dominant decay mode. Here the signal consists of $t \to b\bar b bW$
events against the SM background of $t \to bW$, followed by either
leptonic or hadronic decay of $W$. So one has to look for an excess of
$b$ tags in the $t\bar t$ events compared to the SM prediction
\cite{nine}. With a large number of $t\bar t$ events expected from the
future Tevatron Runs and especially from the LHC one expects to use this
method to extend the $H^\pm$ probe to significantly higher values of
$M_{H^\pm}$ at low $\tan \b (\lsim 1)$.

\subsection{$H^\pm$ Search in the $\t\n$ Channel $(\tan\b > 1)$} 

As discussed earlier the $\tan\b > 1$ region is theoretically
favoured. Fig. 2 shows that $H^\pm \to \t\n$ is the dominant decay
mode over this region. Therefore the $\t$ channel is the most
important channel for $H^\pm$ search. The above mentioned
disappearance (indirect) method is equally applicable to this
channel. The resulting exclusion regions from the $D0\!\!\!/$
\cite{twelve} and CDF \cite{thirteen} experiments can be seen on the
right side of Fig. 1. Evidently the disappearance method is not viable
when the signal is $\lsim 10\%$ of the SM background, since this is
the typical uncertainty in the QCD prediction of $\s_{t\bar t}$. This
explains why the resulting exclusion regions cover only extreme values
of $\tan\b$ (compare Fig. 2). In order to extend the probe to the
theoretically favoured range of $\tan\b$ = 1-50, one has to directly
search for the $t\to b\t\n$ events. Using the universality of $W$
coupling one can easily predict the number of $t \to b \t\n$ events
via $W$ from that of $t \to b \ell \n$ events. Since $H^\pm$ couples
only to the former any excess of $\t$ events over the universality
prediction constitutes a signal for $t \to bH^\pm$ decay. The CDF
group has used a small data sample in the inclusive $\t$ channel to
search for the direct $t \to bH^\pm$ signal \cite{fourteen}. The
resulting exclusion region can be seen on the right side of Fig. 1,
which is roughly overlapping with that obtained via the indirect
method. With the much higher event rates expected from future Tevatron
Runs and LHC it will be better to use the $\ell \t$ channel for
$H^\pm$ search instead of the inclusive $\t$,since the former is a
cleaner and far more robust channel \cite{fifteen,sixteen}. It
corresponds to the decay of one of the $t\bar t$ pair into $\ell$ via
$W$ while the other decays into a $\t$ channel.

\subsection{$\t$ Polarization Effect} 

The discovery reach of the $\t$ channel for $H^\pm$ search at Tevatron
and LHC can be significantly enhanced by exploiting the opposite
polarization of $\t$ coming from the $H^\pm \to \t\n (P_\t = +1)$ and
$W^\pm \to \t\n (P_\t = -1)$ decays \cite{seventeen}. Let us briefly
describe this simple but very powerful method. The best channel for
$\t$-detection in terms of efficiency and purity is its 1-prong
hadronic decay channel, which accounts for 50\% of its total decay
width. The main contributors to this channel are 
\beq
\t^\pm &\to& \p^\pm \n_\t (12.5\%), \ \ \t^\pm \to \r^\pm \n_\t \to 
\p^\pm \p^0 \n_\t (26\%), \nonumber \\ [2mm]
\t^\pm &\to& a^\pm_1 \nu_\t \to \p^\pm \p^0 \p^0 \n_\t (7.5\%),
\label{eq17}
\eeq 
where the branching fractions of the $\p$ and $\r$ channels include
the small $K$ and $K^\ast$ contributions respectively \cite{two},
which have identical polarization effects. Together they account for
more than 90\% of the 1-prong hadronic decay of $\t$. The CM angular
distributions of $\t$ decay into $\p$ or a vector meson $v (= \r,
a_1)$ is simply given in terms of its polarization as
\beq
{1\over \G_\p} \ {d\G_\p \over d\cos\theta} &=& {1\over 2} (1 + P_\t
\cos\theta), \nonumber \\ [2mm]
{1\over \G_v} \ {d\G_{vL} \over d\cos\theta} &=& {{1\over 2} m^2_\t
\over m^2_\t + 2 m^2_v} (1 + P_\t \cos\theta) , \nonumber \\ [2mm]
{1\over \G_v} \ {d\G_{vT} \over d\cos\theta} &=& {m^2_v \over m^2_\t +
2 m^2_v} (1 - P_\t \cos\theta) ,
\label{eq18}
\eeq
where $L,T$ denote the longitudinal and transverse polarization states
of the vector meson \cite{seventeen,eighteen}. This angle is related
to the fraction $x$ of the $\t$ lab. momentum carried by the meson,
i.e. the (visible) $\t$-jet momentum, via
\be
\cos\theta = {2x - 1 - m^2_{\p,v}/m^2_\t \over 1 - m^2_{\p,v}/m^2_\t} .
\label{eq19}
\ee
It is clear from (\ref{eq18}) and (\ref{eq19}) that the signal $(P_\t
= + 1)$ has a harder $\t$-jet than the background $(P_\t = -1)$ for
the $\p$ and the $\r_L, a_{1L}$ contributions; but it is the opposite
for $\r_T, a_{1T}$ contributions. Now, it is possible to suppress the
transverse $\r$ and $a_1$ contributions and enhance the hardness of
the signal $\t$-jet relative to the background even without
identifying the individual resonance contributions to this
channel. This is because the transverse $\r$ and $a_1$ decays favour
even sharing of momentum among the decay pions, while the longitudinal
$\r$ and $a_1$ decays favour uneven distributions, where the charged
pion carries either very little or most of the momentum
\cite{seventeen,eighteen}. Figure 3 shows the decay distributions of
$\r_L, a_{1L}$ and $\r_T, a_{1T}$ in the momentum fraction carried by
the charged pion, i.e.
\be
x' = p_{\p^\pm}/p_{\t {\rm -jet}} .
\label{eq20}
\ee
The distributions are clearly peaked near $x' \simeq 0$ and $x' \simeq
1$ for the longitudinal $\r$ and $a_1$, while they are peaked in the
middle for the transverse ones. Note that the $\t^+ \to \p^\pm \n_\t$
decay would appear as a $\d$ function at $x' = 1$ on this plot. Thus
requiring the $\p^\pm$ to carry $>$ 80\% of the $\t$-jet momentum,
\be
x' > 0.8 ,
\label{eq21}
\ee
retains about half the longitudinal $\r$ along with the pion but very
little of the transverse contributions. This cut suppresses not only
the $W \to \t\n$ background but also the fake $\t$ background from QCD
jets\footnote{Note that the $x'\simeq 0$ peak from $\r_L$ and $a_{1L}$ can not
be used in practice, since $\t$-identification requires a hard
$\p^\pm$, which will not be swept away from the accompanying neutrals
by the magnetic field.}. Consequently the $\t$-channel can be used for
$H^\pm$ search over a wider range of parameters. The resulting $H^\pm$
discovery reach of LHC is shown on the left side of Fig.4
\cite{nineteen}. It goes upto $M_A \simeq 100$ GeV ($M_{H^\pm} \simeq$
130 GeV) around the dip region of $\tan \b \simeq 7.5$ and upto $M_A
\simeq 140$ GeV ($M_{H^\pm} \simeq 160$ GeV) outside this region.

\section{Search for a Heavy $H^\pm (M_{H^\pm} > m_t)$ at LHC}

The main production process here is the leading order (LO) process
\cite{twenty}
\be
gb \to tH^- + h.c. 
\label{eq22}
\ee
The complete NLO QCD corrections have been recently calculated by two groups \cite{twentyone,twentytwo}, in agreement with one another. Their main results are summarized below:
\begin{enumerate}
\item[{(i)}] 
The effect of NLO corrections can be incorporated by multiplying the
above LO cross-section by a $K$ factor, with practically no change in
its kinematic distributions.
\item[{(ii)}] 
With the usual choice of renormalization and factorization scales,
$\m_R = \m_F = M_{H^\pm} + m_t$, one gets $K \simeq 1.5$ over the
large $M_{H^\pm}$ and $\tan\b$ range of interest.
\item[{(iii)}] 
The overall NLO correction of 50\% comes from two main sources --- (a)
$\sim$ 80\% correction from gluon emission and virtual gluon exchange
contributions to the LO process (\ref{eq22}), and (b) $\sim -$ 30\%
correction from the NLO process
\be
gg \to t H^- b + h.c. ,
\label{eq23}
\ee
after subtracting the overlapping piece from (\ref{eq22}) to avoid
double counting.
\item[{(iv)}] 
As clearly shown in \cite{twentytwo}, the negative correction from (b)
is an artifact of the common choice of factorization and
renormalization scales. With a more appropriate choice of the
factorization scale, $\m_F \simeq (M_{H^\pm} + m_t)/5$, the correction
from (b) practically vanishes while that from (a) reduces to $\sim$
60\%. Note however that the overall $K$ factor is insensitive to this
scale variation.
\item[{(v)}] 
Hence for simplicity one can keep a common scale of 
$\m_{F,R} = M_{H^\pm} + m_t$ along with a $K$ factor of 1.5, with an estimated
uncertainty of 20\%. Note that for the process (\ref{eq22}) the
running quark masses of the $H^+ \bar tb$ coupling (\ref{eq10}) are to
be evaluated at $\m_R$, while the patron densities are evaluated at
$\m_F$.
\end{enumerate}

The dominant decay mode for a heavy $H^\pm$ is into the $tb$
channel. The $H^\pm \to \t\n$ is the largest subdominant channel at
large $\tan \b (\gsim 10)$, while the $H^\pm \to W^\pm h^0$ can be the
largest subdominant channel over a part of the small $\tan \b$ region
\cite{one}. Let us look at the prospects of a heavy $H^\pm$ search at
LHC in each of these channels. The dominant background in each case
comes from the $t\bar t$ production process (\ref{eq13}).

\subsection{Heavy $H^\pm$ Search in the $\t\n$ Channel} 

This constitutes the most important channel for a heavy $H^\pm$ search
at LHC in the large $\tan\b$ region. Over a large part of this region,
$\tan \b \gsim 10$ and $M_{H^\pm} \gsim 300$ GeV, we have
\be
BR (H^\pm \to \t\n) = 20 \pm 5 \% .
\label{eq24}
\ee
The $H^\pm$ signal coming from (\ref{eq22}) and (\ref{eq24}) is
distinguished by very hard $\t$-jet and missing-$p_T (p\!\!\!/_T)$,
\be
p^T_{\t{\rm -jet}} > 100 GeV \ {\rm and} \ p\!\!\!/_T > 100 GeV ,
\label{eq25}
\ee
with hadronic decay of the accompanying top quark $(t \to b q \bar q)$
\cite{twentythree}. The main background comes from the $t\bar t$
production process (\ref{eq13}), followed by $t \to b\t\n$, while the
other $t$ decays hadronically. This has however a much softer $\t$-jet
and can be suppressed significantly with the cut
(\ref{eq25}). Moreover the opposite $\t$ polarizations for the signal
and background can be used to suppress the background further, as
discussed earlier. Figure 5 shows the signal and background
cross-sections against the fractional $\t$-jet momentum carried by the
charged pion (\ref{eq20}). The hard charged pion cut of (\ref{eq21})
suppresses the background by a factor of 5-6 while retaining almost
half the signal cross-section. Moreover the signal $\t$-jet has a
considerably harder $p_T$ and larger azimuthal opening angle with the
$p\!\!\!/_T$ in comparison with the background. Consequently the
signal has a much broader distribution in the transverse mass of the
$\t$-jet with the $p\!\!\!/_T$, extending upto $M_{H^\pm}$, while the
background goes only upto $M_W$. Figure 6 shows these distributions
both with and without the hard charged pion cut (\ref{eq21}). One can
effectively separate the $H^\pm$ signal from the background and
estimate the $H^\pm$ mass from this distribution. The LHC discovery
reach of this channel is shown in Fig. 4, which clearly shows it to be
the best channel for a heavy $H^\pm$ search at large $\tan\b$. It
should be added here that the transition region between $M_{H^\pm} >
m_t$ and $< m_t$ has been recently analysed in \cite{twentyfour} by
combining the production process of (\ref{eq22}) with
(\ref{eq13},\ref{eq14}). As a result it has been possible to bridge the gap
between the two discovery contours of Fig. 4 via the $\t\n$ channel.

\subsection{Heavy $H^\pm$ Search in the $tb$ Channel} 

Let us discuss this first for 3 and then 4 b-tags. In the first case
the signal comes from (\ref{eq22}), followed by
\be
H^\pm \to t \bar b, \ \bar t b .
\label{eq26}
\ee
The background comes from the NLO QCD processes 
\be
gg \to t \bar t b \bar b, \ gb \to t \bar t b + h.c., \ gg \to t\bar t g ,
\label{eq27}
\ee
where the gluon jet in the last case can be mistagged as $b$ (with
a typical probability of $\sim 1\%$). One requires leptonic decay of
one of the $t\bar t$ pair and hadronic decay of the other with a $p_T
> 30$ GeV cut on all the jets \cite{twentyfive}. For this cut the
$b$-tagging efficiency at LHC is expected to be $\sim$ 50\%. After
reconstruction of both the top masses, the remaining (3rd) $b$ quark
jet is expected to be hard for the signal (\ref{eq22},\ref{eq26}), but
soft for the background processes (\ref{eq27}). A $\ p_T > 80$ GeV cut
on this $b$-jet improves the signal/background ratio. Finally this
$b$-jet is combined with each of the reconstructed top pair to give
two entries of $M_{tb}$ per event. For the signal events, one of them
corresponds to the $H^\pm$ mass while the other constitutes a
combinatorial background. Figure 7 shows this invariant mass
distribution for the signal along with the above mentioned background
processes for different $H^\pm$ masses at $\tan \b = 40$ Similar
results hold for $\tan \b \simeq 1.5$. One can check that the
significance level of the signal is $S/\sqrt{B} \gsim 5$ \cite{twentyfive}. The
corresponding $H^\pm$ discovery reaches in the high and low $\tan \b$
regions are shown in Fig. 4. While the discovery reach via $tb$ is
weaker than that via the $\t\n$ channel in the high $\tan \b$ region,
the former offers the best $H^\pm$ discovery reach in the low $\tan
\b$ region. This is particularly important in view of the fact that
the indirect LEP limit shown in Fig. 4 gets significantly weaker with
the reported increase in the top quark mass, as discussed
earlier. Indeed this $H^\pm \to tb$ discovery contour constitutes the
most robust discovery limit for the MSSM Higgs sector over the low
$\tan\b$ region. On the other hand the $H^\pm \to \t\n$ contour is
competitive with that from the $H^0/A^0 \to \t\t$ channel as the best
MSSM Higgs discovery limit over the high $\tan\b$ region. Finally the
corresponding $H^\pm \to \t\n$ contour from $t \to bH^+$ decay, also
shown in Fig. 4, constitutes the best discovery limit of the MSSM
Higgs sector over the low $M_A$ region (see e.g. Fig. 27 of
ref.\cite{one}).
 
One can also use 4 $b$-tags to look for the $H^\pm \to tb$ signal
\cite{twentysix}. The signal comes from (\ref{eq23},\ref{eq26}), and the
background from the first process of (\ref{eq27}). After the
reconstruction of the $t\bar t$ pair, both the remaining pair of
$b$-jets are expected to be soft for the background, since they come
from gluon splitting. For the signal, however, one of them comes from
the $H^\pm$ decay (\ref{eq26}); and hence expected to be hard and
uncorrelated with the other $b$-jet. Thus requiring a $p_T >$ 120 GeV
cut on the harder of the two $b$-jets along with large invariant mass
$(M_{bb} > 120$ GeV) and opening angle (cos$\theta_{bb} < 0.75$) for the
pair, one can enhance the signal/background ratio
substantially. Unfortunately the requirement of 4 $b$-tags makes the
signal size very small. Moreover the signal contains one soft $b$-jet
from (\ref{eq23}), for which one has to reduce the $p_T$ threshold from
30 to 20 GeV. The resulting signal and background cross-sections are
shown in Fig. 8 for $\tan \b = 40$. In comparison with Fig. 7 one can
see a significant enhancement in the signal/background ratio, but at
the cost of a much smaller signal size. Nonetheless this can be used
as a supplementary channel for $H^\pm$ search, provided one can
achieve good $b$-tagging for $p_T \sim 20$ GeV jets.

\subsection{Heavy $H^\pm$ Search in the $Wh^0$ Channel}

The tree level coupling for this channel is 
\be
H^+W^-h^0 \colon {1\over 2} g \cos (\b - \a) q_h ,
\label{eq28}
\ee
where $q_h$ is the $h^0$ momentum in the $H^+$ rest frame. The LEP
limit of $M_{h^0} \gsim 100 $ GeV in the MSSM implies that the $H^\pm
\to Wh^0$ decay channel has at least as high a threshold as the $tb$
channel. The maximum value of its decay BR,
\be
B^{\rm max} (H^\pm \to Wh^0) \simeq 5 \% ,
\label{eq29}
\ee
is reached for $H^\pm$ mass near this threshold and low $\tan\b$. The
small BR for this decay channel is due the suppression of the
$H^+W^-h^0$ coupling (\ref{eq28}) by the $q_h$ and the $\cos (\b -
\a)$ factors relative to the $H^+ \bar t b$ coupling
(\ref{eq10}). Note that both the decay channels correspond the same
final state, $H^\pm \to b\bar b W$, along with an accompanying top
from the production process (\ref{eq22}). Nonetheless one can
distinguish the $H^\pm \to Wh^0$ from the $H^\pm \to tb$ as well as
the corresponding backgrounds (\ref{eq27}) by looking for a clustering
of the $b\bar b$ invariant mass around $M_{h^0}$ along with a veto on
the second top \cite{twentyseven}. Unfortunately the BR of
(\ref{eq29}) is too small to give a viable signal for this decay
channel. Note however that the LEP limit of $M_{h^0} \gsim 100$ GeV does
not hold in the CP violating MSSM \cite{five} or the singlet
extensions of the MSSM Higgs sector like the NMSSM
\cite{six}. Therefore it is possible to have a $Wh^0$ threshold
significantly below $m_t$ in these model. Consequently one can have a
$H^\pm$ boson lighter than the top quark in these models in the low
$\tan \b$ region, which can dominantly decay into the $Wh^0$
channel. Thus it is possible to have spectacular $t \to bH^+ \to
bWh^0$ decay signals at LHC in the NMSSM \cite{twentyseven} as well as
the CP violating MSSM \cite{twentyeight}.

\section{Concluding Remarks}

Let me conclude by commenting on a few aspects of $H^\pm$ boson
search, which could not be discussed in this brief review. The
associated production of $H^\pm$ with $W$ boson has been investigated
in \cite{twentynine}, and the $H^\pm H^\mp$ and $H^\pm A^0$
productions in \cite{thirty}. Being second order electroweak
processes, however, they give much smaller signals than (\ref{eq22}),
while suffering from the same background. However one can get
potentially large $H^\pm$ signal from the decay of strongly produced
squarks and gluinos at LHC, which can help to fill in the gap in the
intermediate $\tan\b$ region of Fig. 4 for favourable SUSY parameters
\cite{thirtyone}.

Finally, the virtual SUSY contribution to the NLO correction for
$H^\pm$ production can be potentially important since it is known to
be nondecoupling, i.e. it remains finite even for very large SUSY mass
parameters. The reason for this of course is that the $H^\pm$ mass is
related to the superparticle masses in SUSY models -- e.g. in minimal
SUGRA model the $H^\pm$ mass is of similar size as the sfermion
masses. Therefore the two mass scales can not be
decoupled. Consequently the calculation of virtual SUSY correction to
$H^\pm$ production has received a lot of attention \cite{thirtytwo,
thirtythree, thirtyfour}. The main contribution comes from the virtual
squark-gluino exchange contribution to the $H^+ \bar t b$ vertex. Its
effect can be approximated by a renormalisation of the the $m_{b,t}$
factors in the corresponding coupling (\ref{eq10}) by $1/(1 +
\D_{b,t})$, where \cite{one, twentytwo}
\beq
\D_b &\simeq& {2\a_s \over 3\p} m_{\tilde g} (-A_b + \m\tan\b) I 
(m_{\tilde b 1}, m_{\tilde b 2}, m_{\tilde g}) , \nonumber \\ [2mm] 
I(a,b,c) &=& - {a^2b^2 \ell n (a^2/b^2) + b^2c^2 \ell n (b^2/c^2) +
c^2a^2 \ell n (c^2/a^2) \over (a^2-b^2) (b^2-c^2) (c^2-a^2)} \nonumber
\\ [2mm] && \quad \sim 1/{\rm max} (a^2, b^2, c^2) ;
\label{eq30}
\eeq
and there is a similar expression for $\D_t$. Thus in the large
$\tan\b$ region, where the $m_b$ term dominates the $H^+ \bar t b$
coupling, and for $m_{\tilde g} \gg m_{\tilde b1,2}$, we get
\be
\D_b \sim {2\a_s \over 3\p} \ {\m \tan \b \over m_{\tilde g}} ,
\label{eq31}
\ee
which can be very large for $|\m| \gg m_{\tilde g}$
\cite{thirtyfour}. On the other hand in most SUSY models of common
interest we have $|\m| \sim M_Z$ for naturalness, while $m_{\tilde g}
\gg M_Z$. Therefore the above SUSY correction has only modest effect
on $H^\pm$ production in these models. Indeed a systematic study of
this effect for the `snowmass points and slopes \cite{thirtyfive}',
carried out in \cite{twentytwo}, shows that the SUSY correction to the
cross-section for the LO process (\ref{eq22}) remains $\lsim 20\%$ for
$\tan\b \lsim 30$. This is true not only for minimal SUGRA but for
other popular alternatives like gauge and anomaly mediated SUSY
breaking models as well. As mentioned earlier, the theoretical
uncertainly in the estimate of the NLO QCD correction ($K$ factor) in
the SM is also $\sim 20\%$ \cite{twentytwo}. Therefore one need not
worry too much about the effect of SUSY quantum correction on the
$H^\pm$ boson signal at LHC.

\newpage

\begin{figure*}[t]
\begin{center}
\epsfig{file=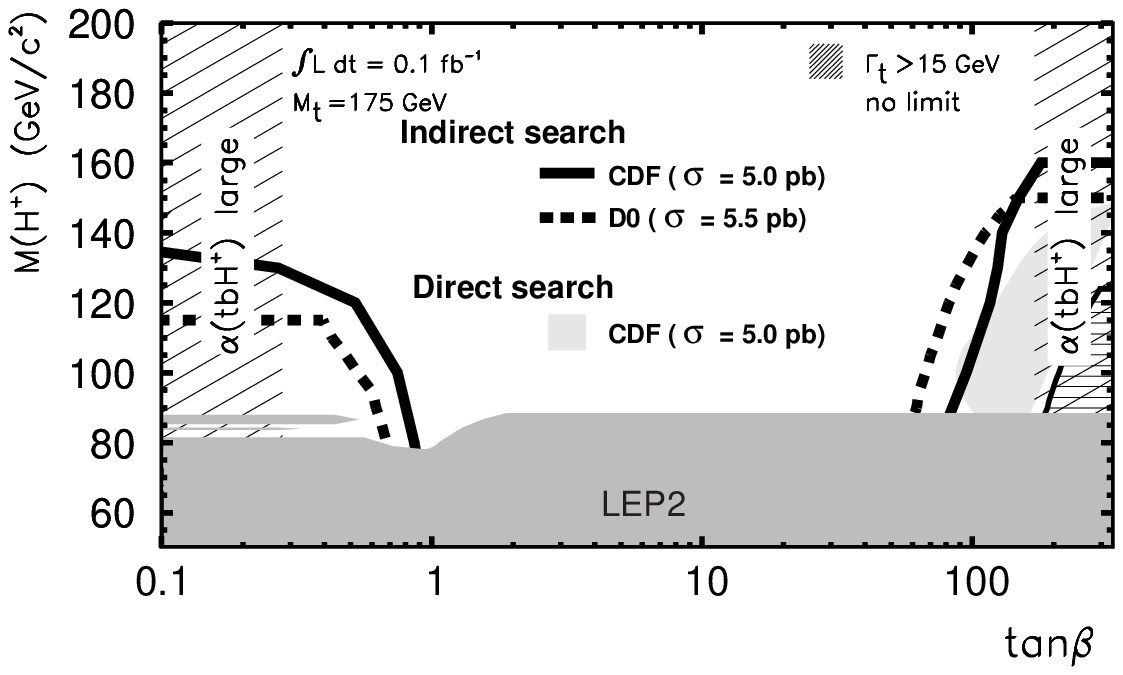,height=8cm,width=8cm,angle=0}
\end{center}
\caption{\label{fig1} 
The 95\% CL limits in $M_{H^\pm} - \tan\b$ plane from LEP-2 (dark grey
band) and Tevatron \cite{two}. The Tevatron indirect search limits
from $D0\!\!\!/$ \cite{twelve} and CDF \cite{thirteen} experiments are
shown along with the direct search limit from CDF \cite{fourteen}. The
cross-hatched regions at extreme values of $\tan\b$ lie outside the
perturbative bounds of eq.(\ref{eq12}).}
\end{figure*}

\begin{figure*}[hb]
\begin{center}
\leavevmode
\epsfig{file=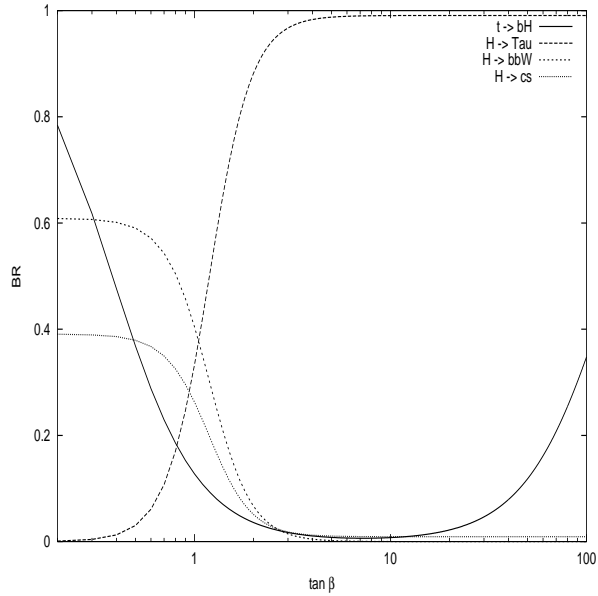,height=8cm,width=8cm,angle=-90}
\end{center}
\caption{\label{fig2} 
The Branching Ratio of top decay into a 140 GeV $H^\pm$ boson
(\ref{eq14}) shown against $\tan\b$ along with those for the three
main decay modes (\ref{eq15}) of this $H^\pm$ boson.}
\end{figure*}

\begin{figure*}[ht]
\begin{center}
\leavevmode
\epsfig{file=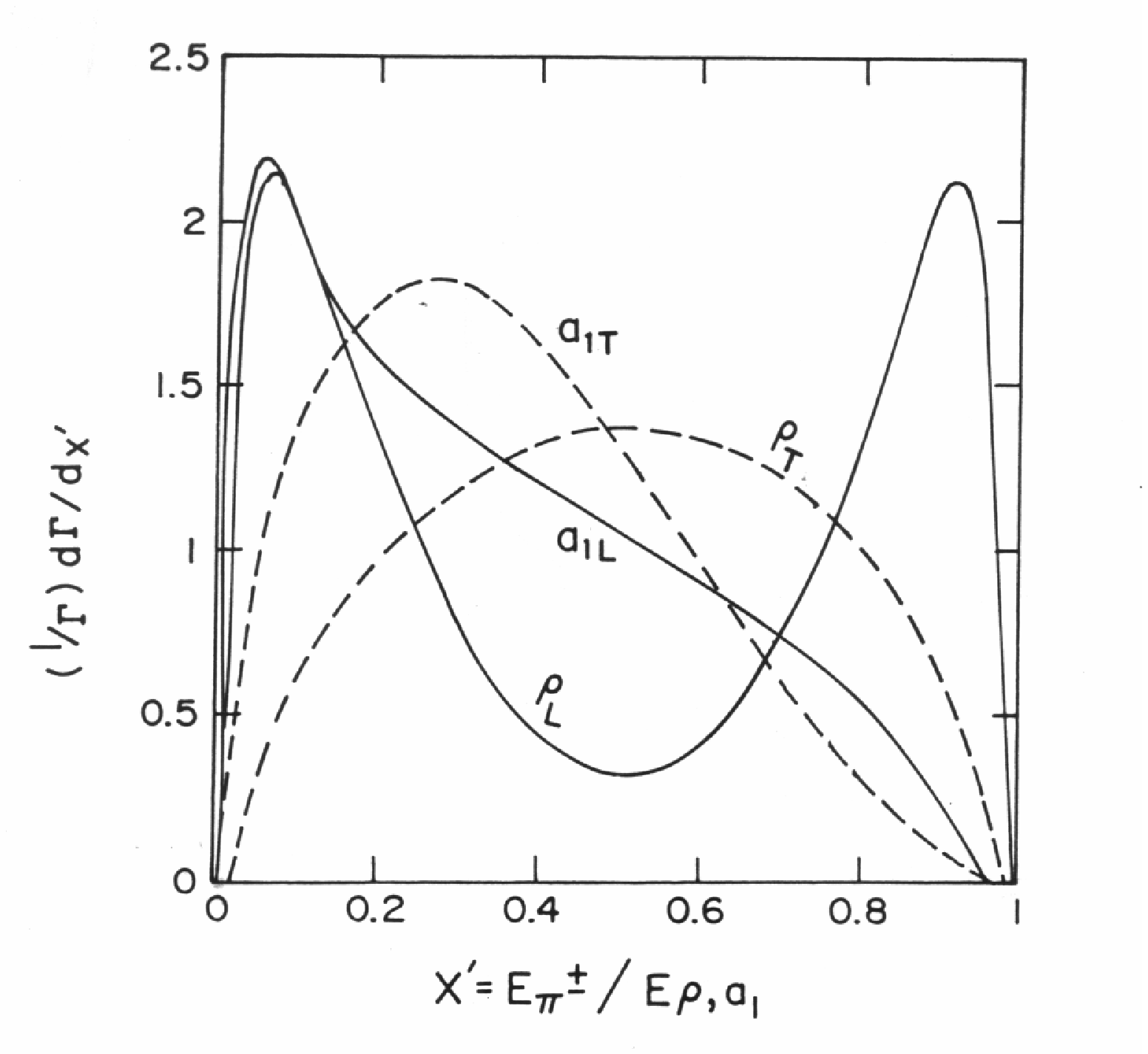,height=8cm,width=8cm,angle=0}
\end{center}
\caption{\label{fig3} 
Distributions of the normalised decay widths of $\t^\pm$ via
$\r^\pm_{L,T} \to \p^\pm \p^0$ and $a^\pm_{1L,T} \to \p^\pm \p^0 \p^0$
in the momentum fraction carried by the charged pion
\cite{seventeen}. On this plot the $\t^\pm \to \p^\pm \n$ decay would
correspond to a $\delta$-function at $x' = 1$. }
\end{figure*}

\begin{figure*}[hb]
\begin{center}
\leavevmode
\epsfig{file=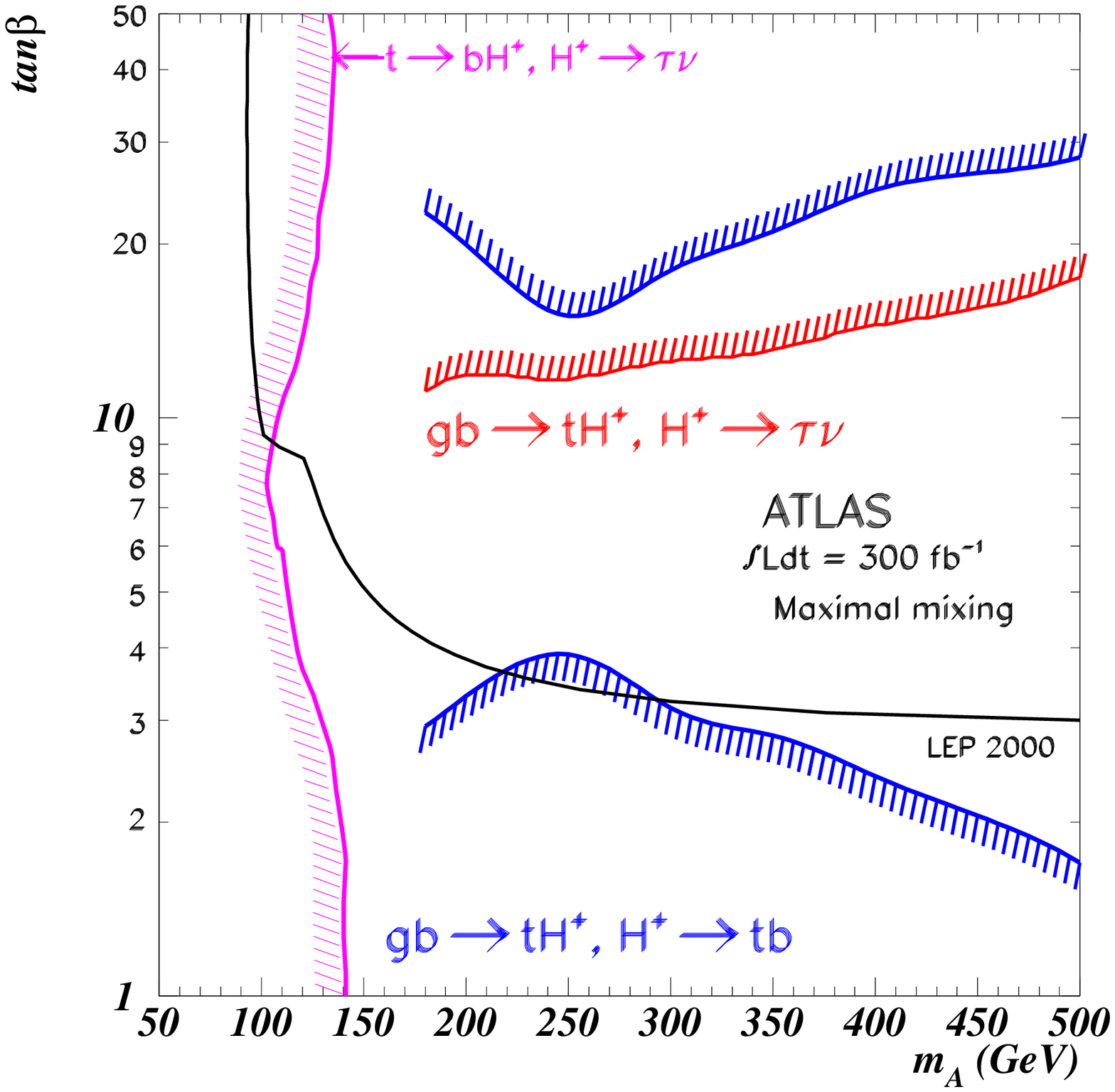,height=8cm,width=8cm,angle=0}
\end{center}
\caption{\label{fig4} 
The 5-$\s$ $H^\pm$ boson discovery contours of the ATLAS experiment at
LHS from $t \to bH^+, H^+ \to \t\n$ (vertical); $gb \to tH^-, H^-
\t\n$ (middle horizontal) and $gb \to tH^-, H^- \to \bar tb$ (upper
and lower horizontal) channels \cite{nineteen}. One can see similar
contours for the CMS experiment in the second paper of
ref.\cite{nineteen}. The horizontal part of indirect LEP limit shown
here has weakened significantly now as explained in the text. }
\end{figure*}

\begin{figure*}[ht]
\begin{center}
\leavevmode
\epsfig{file=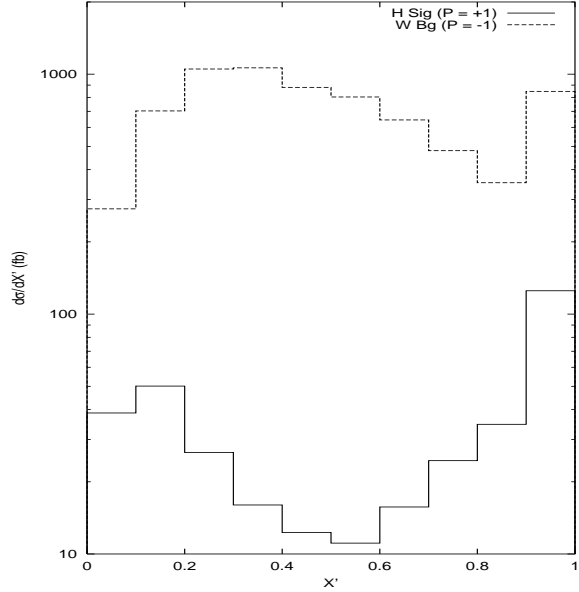,height=8cm,width=8cm,angle=0}
\end{center}
\caption{\label{fig5} 
The LHC cross-section for a 300 GeV $H^\pm$ signal at $\tan\b = 40$ 
shown along with the $t\bar t$ background in the 1-prong $\t$-jet channel,
as functions of the $\t$-jet momentum fraction carried by the charged pion.}
\end{figure*}

\begin{figure*}[hb]
\begin{center}
\leavevmode
\epsfig{file=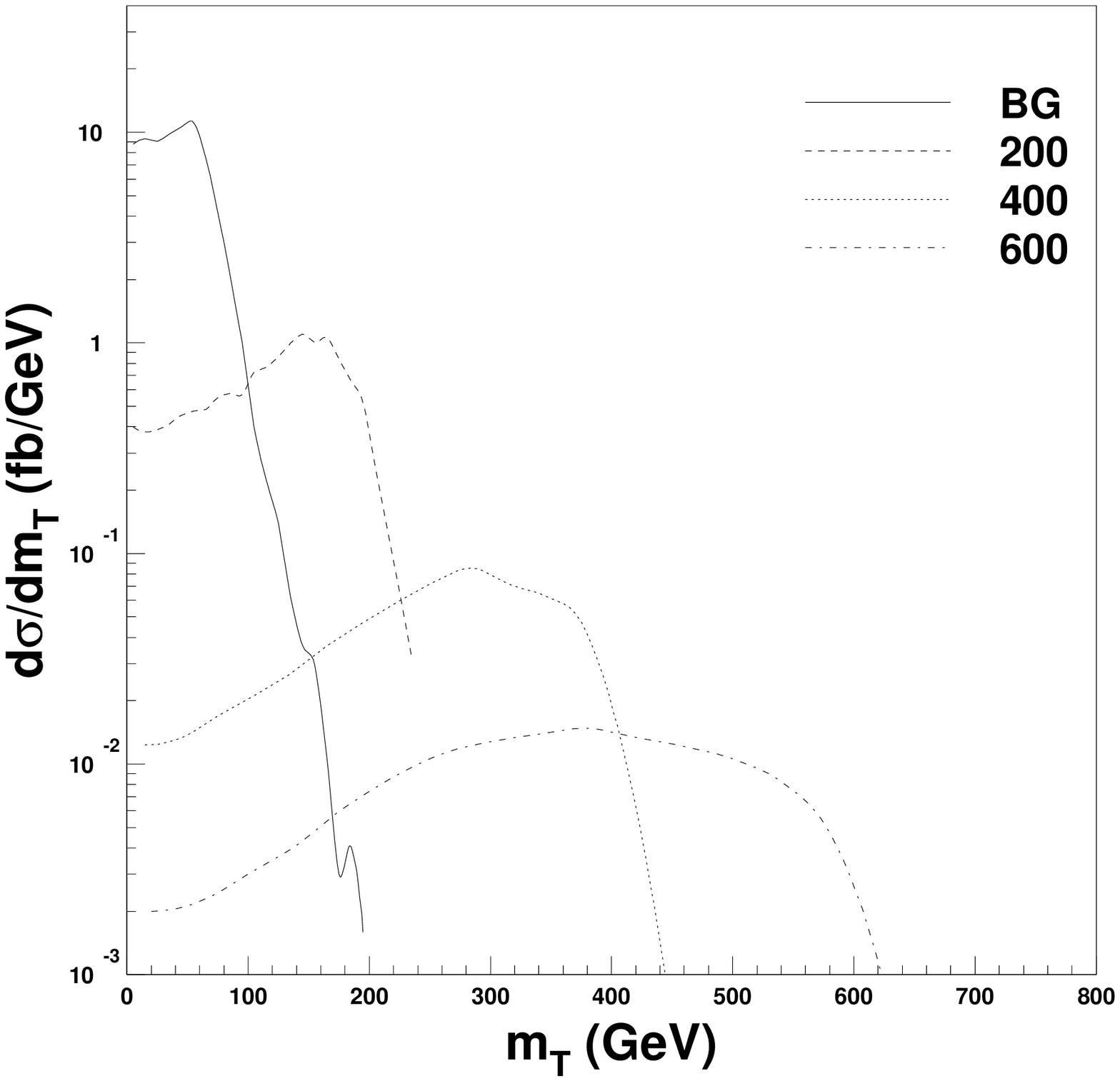,height=8cm,width=8cm,angle=0}
\epsfig{file=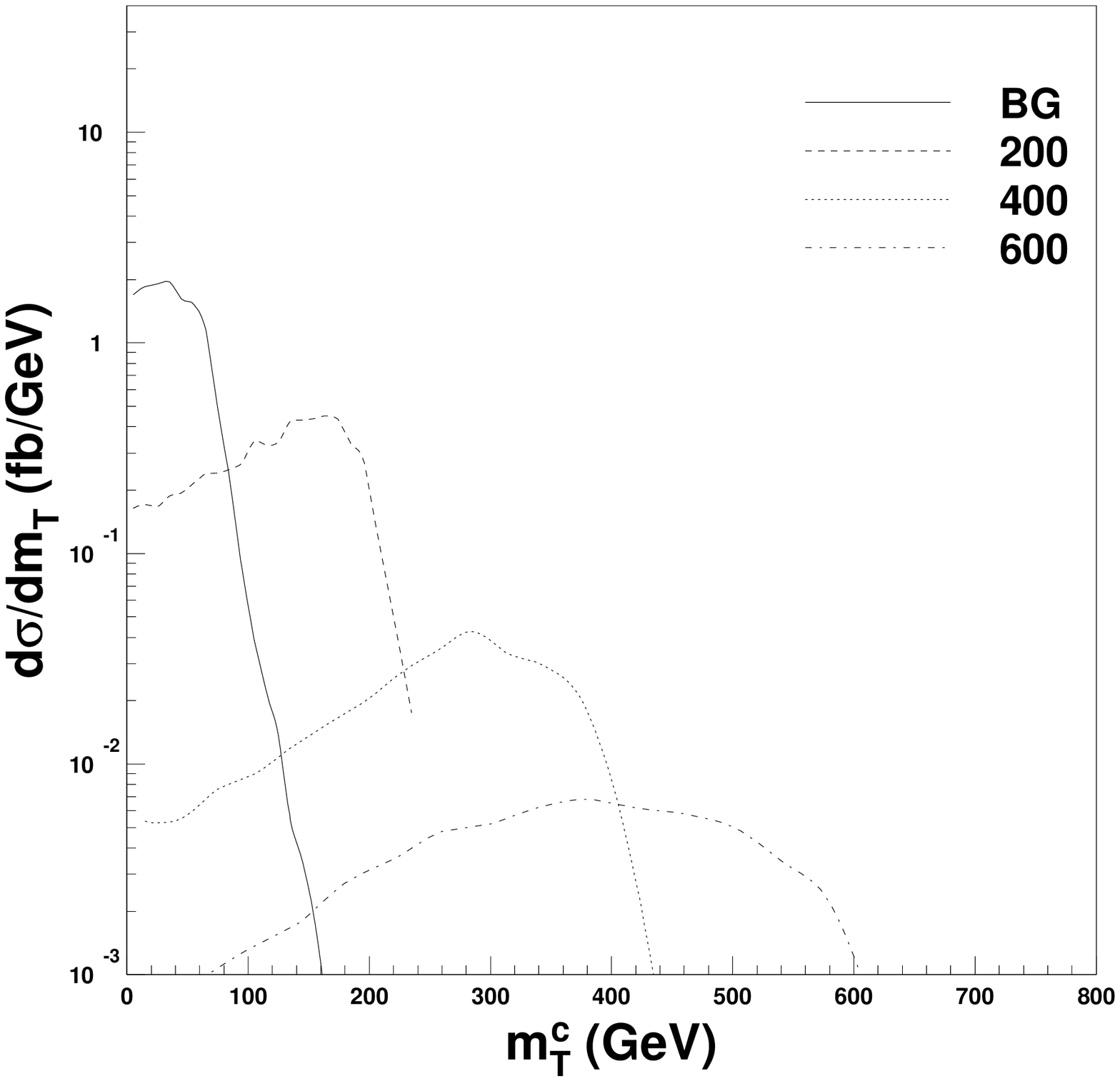,height=8cm,width=8cm,angle=0}
\end{center}
\caption{\label{fig6} 
Distributions of the $H^+$ signal and the $t\bar t$ background
cross-sections in the transverse mass of the $\t$-jet with
$p\!\!\!/_T$ for (left) all 1-prong $\t$-jets, and (right) those with
the charged pion carrying $> 80\%$ of the $\t$-jet momentum
($M_{H^\pm}$ = 200,400,600 GeV and $\tan\b$ = 40) \cite{twentythree}.}
\end{figure*}

\begin{figure*}[ht]
\begin{center}
\leavevmode
\epsfig{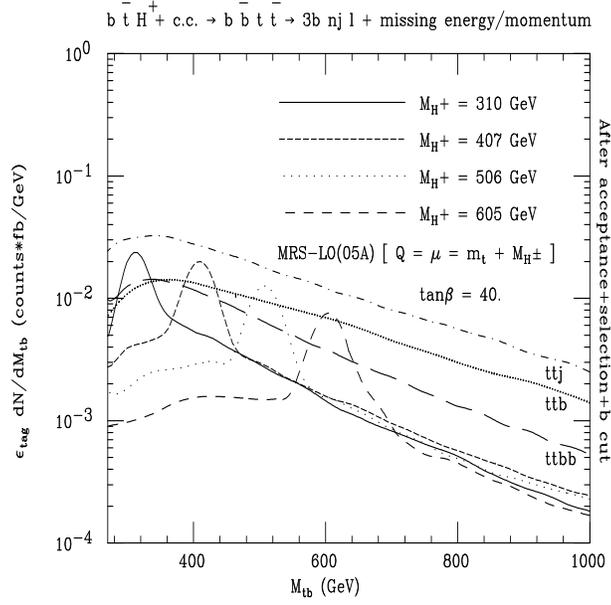}
\end{center}
\caption{\label{fig7} 
The reconstructed $tb$ invariant mass distribution of the $H^\pm$
signal and different QCD backgrounds in the isolated lepton plus
multijet channel with 3 $b$-tags \cite{twentyfive}.}
\end{figure*}

\begin{figure*}[hb]
\begin{center}
\leavevmode
\epsfig{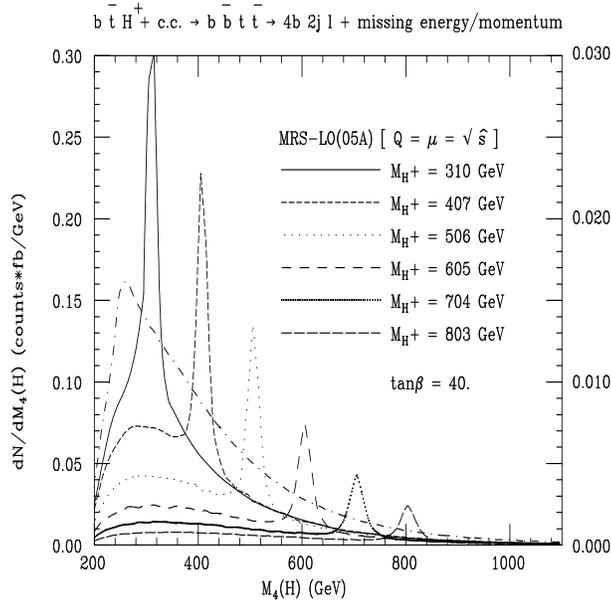}
\end{center}
\caption{\label{fig8} 
The reconstructed $tb$ invariant mass distribution of the $H^\pm$
signal and the QCD background in the isolated lepton plus multijet
channel with 4 $b$-tags \cite{twentysix}. The scale on the right
corresponds to applying a $b$-tagging efficiency factor $\e^4_b =
0.1$.}
\end{figure*}


\begin{thebibliography}{99}
\bibitem{one} 
For a recent review see M. Carena and H. Haber,
{\it Prog. Part. Nucl. Phys.} {\bf 50}, 63 (2003) [hep-ph/0208209].

\bibitem{two}
K. Hagiwara et al. (Particle Data Group), {\it Phys. Rev.} {\bf D66},
010001 (2002) (URL: http://pdg.lbl.gov)

\bibitem{three}
CDF and D0 Collaborations and Tevatron Electroweak Working Group,
hep-ex/0404010.

\bibitem{four}
G. Degrassi, S. Hainemeyer, W. Hollik, P. Slavich and G. Weiglein,
{\it Euro. Phys. J.} {\bf C28}, 133 (2003).

\bibitem{five}
M. Carena, J. Ellis, S. Mrenna, A. Pilaftsis and C.E.M. Wagner, {\it
Nucl. Phys.} {\bf B659}, 145 (2003).

\bibitem{six}
M. Drees, E. Ma, P.N. Pandita, D.P. Roy and S. Vempati, {\it Phys. Lett.} 
{\bf B433}, 346 (1998); see also C. Panagiotakopoulos and A. Pilftsis,
{\it Phys. Lett.} {\bf B505}, 184 (2001).

\bibitem{seven}
A. Mendez and A. Pomarol, {\it Phys. Lett.} {\bf B252}, 461 (1990); 
C.S. Li and R.J. Oakes, {\it Phys. Rev.} {\bf D43}, 855 (1991); M. Drees 
and D.P. Roy, {\it Phys. Lett.} {\bf B269}, 155 (1991). 

\bibitem{eight}
S. Moretti and W.J. Stirling, {\it Phys. Lett.} {\bf B347}, 291 (1995); 
{\bf B366}, 451 (E) (1996); A. Djouadi, J. Kalinowski and P.M. Zerwas, 
{\it Z. Phys.} {\bf C70}, 435 (1996). 

\bibitem{nine}
E. Ma, D.P. Roy and J. Wudka, {\it Phys. Rev. Lett.} {\bf 80}, 1162 (1998). 

\bibitem{ten}
E. Keith, E. Ma and D.P. Roy, {\it Phys. Rev.} {\bf D56}, R5306 (1997). 

\bibitem{eleven}
S. Katani et al., {\it Phys. Lett.} {\bf B378}, 329 (1996); E.L. Berger and 
H. Contopanagos, {\it Phys. Rev.} {\bf D54}, 3085 (1996). 

\bibitem{twelve}
$D0\!\!\!/$ Collaboration: {\it Phys. Rev. Lett.} {\bf 82}, 4975 (1999). 
 
\bibitem{thirteen}
CDF Collaboration: {\it Phys. Rev.} {\bf D62}, 012004 (2000). 

\bibitem{fourteen}
CDF Collaboration: {\it Phys. Rev. Lett.} {\bf 79}, 357 (1997). 

\bibitem{fifteen}
M. Guchait and D.P. Roy, {\it Phys. Rev.} {\bf D55}, 7263 (1997). 

\bibitem{sixteen}
CDF Collaboration: {\it Phys. Rev.} {\bf D62}, 012004 (2000). 

\bibitem{seventeen}
S. Raychaudhuri and D.P. Roy, {\it Phys. Rev.} {\bf D52}, 1556 (1995); 
{\it Phys. Rev.} {\bf D53}, 4902 (1996). 

\bibitem{eighteen}
B.K. Bullock, H. Hagiwara and A.D. Martin, {\it Nucl. Phys.} {\bf B395}, 499 
(1993). 

\bibitem{nineteen}
K.A. Assamagan, Y. Coadou and A. Deandrea, {\it Eur. Phys. J} {\bf C4}, 9
(2002); see also D. Denegri et al., CMS Note 2001/032, hep-ph/0112045. 

\bibitem{twenty}
A.C. Bawa, C.S. Kim and A.D. Martin, {\it Z. Phys.} {\bf C47}, 75 (1990);
J.F. Gunion, {\it Phys. Lett.} {\bf B322}, 125 (1994); V. Barger, R.J.N. 
Phillips and D.P. Roy, {\it Phys. Lett.} {\bf B324}, 236 (1994).

\bibitem{twentyone}
S.H. Zhu, {\it Phys. Rev.} {\bf D67}, 075006 (2003). 

\bibitem{twentytwo}
T. Plehn, {\it Phys. Rev.} {\bf D67}, 014018 (2003); E.L. Berger, T. Han, 
J. Jiang and T. Plehn, hep-ph/0312286.

\bibitem{twentythree}
D.P. Roy, {\it Phys. Lett.} {\bf B459}, 607 (1999).

\bibitem{twentyfour}
K.A. Assamagan, M. Guchait and S. Moretti, hep-ph/0402057; see also
F. Borzumati, J.L. Kneur and N. Polonsky, {\it Phys. Rev.} {\bf D60},
115011 (1999).

\bibitem{twentyfive}
S. Moretti and D.P. Roy, {\it Phys. Lett.} {\bf B470}, 209 (1999). 

\bibitem{twentysix}
D.J. Miller, S. Moretti, D.P. Roy and W.J. Stirling, {\it Phys. Rev.} 
{\bf D61}, 055011 (2000); see also K.A. Assamagan and N. Gollub,
hep-ph/0406013.

\bibitem{twentyseven}
M. Drees, M. Guchait and D.P. Roy, {\it Phys. Lett.} {\bf B471}, 39 (1999). 

\bibitem{twentyeight}
D.K. Ghosh, R.M. Godbole and D.P. Roy (in preparation). 

\bibitem{twentynine}
A.A. Barrientos Bendezu and B.A. Kniehl, {\it Phys. Rev.} {\bf D59}, 
015009 (1999); S. Moretti and K. Odagiri, {\it Phys. Rev.} {\bf D59}, 
055008 (1999); O. Brein, H. Hollik and S. Kanemura, {\it Phys. Rev.} {\bf D63},
095001 (2001). 

\bibitem{thirty}
A. Kraus, T. Plehn, M. Spira and P.M. Zerwas, {\it Nucl. Phys.} {\bf
B519}, 85 (1998); O. Brein and H. Hollik, {\it Eur. Phys. J.} {\bf
C13}, 175 (2000); A.A. Barrientos Bendezu and B.A. Kniehl, {\it
Nucl. Phys.} {\bf B568}, 305 (2000). For associated $H^\pm A^0$
production see Q.H. Cao, S. Kanemura and C.P. Yuan, {\it Phys. Rev.} 
{\bf D69}, 075008 (2004).

\bibitem{thirtyone}
M. Bisset, M. Guchait and S. Moretti, {\it Eur. Phys. J.} {\bf C19},
143 (2001); A. Datta, A. Djouadi, M. Guchait and Y. Mambrini, {\it
Phys. Rev.} {\bf D65}, 015007 (2002).

\bibitem{thirtytwo}
L. Hall, R. Rattazzi and U. Sarid, {\it Phys. Rev.} {\bf D50}, 7048
(1994); M. Carena, M. Olechowski, S. Pokorski and C.E.M. Wagner, {\it
Nucl. Phys.} {\bf B426}, 269 (1994).

\bibitem{thirtythree}
J.A. Coarasa, R.A. Jimenez and J. Sola, {\it Phys. Lett.} {\bf B389},
312 (1996); R.A. Jimenez and J. Sola, {\it Phys. Lett.} {\bf B389}, 53
(1996); A. Bartl, H. Eberl, K. Hikasa, K. Kon, W. Majerotto and
Y. Yamada, {\it Phys. Lett.} {\bf B378}, 167 (1996).

\bibitem{thirtyfour}
A. Belyaev, D. Garcia, J. Guasch and J. Sola, {\it Phys. Rev.} {\bf D65}, 031701 (2002); {\it JHEP} {\bf 0206}, 059 (2002). 

\bibitem{thirtyfive}
B.C. Allanach et al., in Proc. of the APS/DPF/DPB Summer Study on the
Future of Particle Physics (Snowmass 2001) {\it Eur. Phys. J.} {\bf
C25}, 113 (2002).

\end{thebibliography}
\end{document}